# Flow of Electrically Charged Fluids Through Elastic Porous Media

Jiashi Yang (jyang1@unl.edu)
Department of Mechanical and Materials Engineering
University of Nebraska-Lincoln, Lincoln, NE 68588-0526, USA

**Abstract**
We study the flow of an electrically charged fluid through an elastic and porous medium. A three-continuum model consisting of an elastic solid, a viscous fluid, and a mobile-charge continuum is used. The relevant laws of physics are applied systematically to the constituents or the combined continuum, leading to their continuity equations for conservation of mass or charge, linear and angular momentum equations as well as constitutive relations. The analysis assumes quasistatic electric fields and is for nonmagnetizable materials. The resulting theory is nonlinear, valid for large deformations and strong fields and can be specialized or generalized in various ways.

## 1. Introduction

Flow of a fluid through a porous medium is a classical research area of mechanics with many applications [1]. The fluid and solid involved may also have various electrical behaviors such as polarization or conduction [2,3]. For electrically charged and conducting continua, nonlinear continuum theories can be obtained systematically using models based on motions of charged particles [4] or properly constructed multi-continuum models [5–17] along with systematic applications of the relevant laws of physics. The multi-continuum models can be viewed as mixtures of thermomechanical and electromagnetic constituents [18]. In this paper we study the flow of an electrically charged and conducting fluid through an elastic and porous medium. We use the multi-continuum model which in this case is of moderate complexity and can be either reduced to models for simpler material behaviors or generalized for more complicated material systems.

## 2. Multi-Continuum Model

The model consists of three continua called the solid, fluid and mobile-charge or electron continua, respectively. The fields associated with the three continua are indicated by superscripts $s$, $f$ and $e$, respectively. The electric field $\mathbf{E}$ and the absolute temperature $\theta$ in the three constituents are assumed to be the same which may be viewed as an approximation and can be removed or improved if needed [18]. The solid continuum is assumed to be a dielectric which in fact consists of a lattice continuum and a bound-charge continuum [7]. It polarizes under an electric field but is not electrically conducting by itself. It is assumed to be electrically neutral in its reference state without the so-called residual charges [11]. In this paper, the electrical polarization as described by the lattice and bound-charge continua in [7] is considered as already understood. Hence we treat the solid continuum effectively like one continuum with known electric body force, couple and power due to polarization as calculated from [7]. The fluid continuum is similar. The electron continuum is responsible for electrical conduction. Because of the presence of the fluid and electron continua, we begin with the spatial description. $\mathbf{y}$ or $y_k$ are the present coordinates of a point in space occupied by material particles of the three continua. Let $\rho^s$, $\rho^f$ and $\rho^e$ be the apparent mass densities of the constituents. $\mu^e$ is the apparent charge density of the electron continuum. $\mathbf{v}^s$, $\mathbf{v}^f$ and $\mathbf{v}^e$ are the velocity fields of the corresponding constituents. They are all functions of $\mathbf{y}$ and time $t$. Let $v$ be a fixed region. Its boundary surface is $s$ which has an outward unit normal $\mathbf{n}$. We have the following continuity equations for the conservation of mass or charge of the above constituents:

$$\frac{\partial \rho^s}{\partial t} + (\rho^s v_k^s)_{,k} = 0, \tag{2.1}$$

$$\frac{\partial \rho^f}{\partial t} + (\rho^f v_k^f)_{,k} = 0, \tag{2.2}$$

$$\frac{\partial \rho^e}{\partial t} + (\rho^e v_k^e)_{,k} = 0, \tag{2.3}$$

$$\frac{\partial \mu^e}{\partial t} + (\mu^e v_k^e)_{,k} = 0. \tag{2.4}$$

Equations (2.3) and (2.4) are related by the charge-to-mass ratio. Equation (2.4) may be written as

$$\frac{\partial \mu}{\partial t} + J_{k,k} = 0, \quad \mu = \mu^e, \quad \mathbf{J} = \mu^e \mathbf{v}^e, \tag{2.5}$$

where $\mathbf{J}$ is the current density. The Maxwellian electric field $\mathbf{E}$ is assumed to be quasistatic and satisfies

$$\nabla \times \mathbf{E} = 0, \quad E = -\nabla \varphi, \tag{2.6}$$

$$\varepsilon_0 E_{k,k} = -P_{k,k}^s - P_{k,k}^f + \mu^e, \tag{2.7}$$

where $\varphi$ is the electric potential and $\mathbf{P}$ the electric polarization vector. $\varepsilon_0$ is the permittivity of free space. $-P_{k,k}^s$ and $-P_{k,k}^f$ are the effective polarization charge densities of the solid and fluid, respectively. Equation (2.7) may be written as

$$\nabla \cdot \mathbf{D} = \mu, \tag{2.8}$$

where the electric displacement vector $\mathbf{D}$ is defined by

$$\mathbf{D} = \varepsilon_0 \mathbf{E} + \mathbf{P}, \quad \mathbf{P} = \mathbf{P}^s + \mathbf{P}^f. \tag{2.9}$$

Equations $(2.5)_1$ and (2.8) imply

$$\left(\frac{\partial D_k}{\partial t} + J_k\right)_{,k} = 0. \tag{2.10}$$

### 3. Linear Momentum Equations

The electron continuum is assumed to behave like an ideal fluid mechanically, whose stress field is simply given by $\boldsymbol{\tau}^e = -p^e \mathbf{1}$ as a pressure field where $\mathbf{1}$ is the second-order unit tensor. Let the surface traction per unit area be $\mathbf{t}$, the mechanical body force per unit mass be $\mathbf{f}$, and the electric body force per unit volume be $\mathbf{F}$. They can be associated with various constituents with the corresponding superscripts. The linear momentum equations for individual constituents are

$$\frac{\partial}{\partial t}\int_v \rho^s \mathbf{v}^s dv = -\int_s \rho^s \mathbf{v}^s (\mathbf{v}^s \cdot \mathbf{n}) ds + \int_s \mathbf{t}^s ds \\ + \int_v (\rho^s \mathbf{f}^s + \mathbf{F}^{Es} + \mathbf{F}^{sf} - \mu^e \mathbf{E}^{es}) dv, \quad (3.1)$$

$$\frac{\partial}{\partial t}\int_v \rho^f \mathbf{v}^f dv = -\int_s \rho^f \mathbf{v}^f (\mathbf{v}^f \cdot \mathbf{n}) ds + \int_s \mathbf{t}^f ds \\ + \int_v (\rho^f \mathbf{f}^f + \mathbf{F}^{Ef} + \mathbf{F}^{fs} - \mu^e \mathbf{E}^{ef}) dv, \quad (3.2)$$

$$\frac{\partial}{\partial t}\int_v \rho^e \mathbf{v}^e dv = -\int_s \rho^e \mathbf{v}^e (\mathbf{v}^e \cdot \mathbf{n}) ds + \int_s -p^e \mathbf{n} ds \\ + \int_v (\rho^e \mathbf{f}^e + \mathbf{F}^{Ee} + \mu^e \mathbf{E}^{es} + \mu^e \mathbf{E}^{ef}) dv, \quad (3.3)$$

where the interaction between the electron continuum and the solid is described by an effective electric field $\mathbf{E}^{es}$. Similarly, the interaction between the electron continuum and the fluid is described by $\mathbf{E}^{ef}$. The interaction between the solid and fluid is represented by the following effective body force:

$$\mathbf{F}^{sf} = -\mathbf{F}^{fs}. \quad (3.4)$$

The electrostatic body forces $\mathbf{F}^E$ on various constituents are [7,11]

$$F_j^{Es} = P_i^s E_{j,i}, \quad F_j^{Ef} = P_i^f E_{j,i}, \quad \mathbf{F}^{Ee} = \mu^e \mathbf{E}. \quad (3.5)$$

Using the divergence theorem and conservation of mass, we obtain the differential form of Eq. (3.1) as

$$\nabla \cdot \boldsymbol{\tau}^s + \rho^s \mathbf{f}^s + \mathbf{F}^{Es} + \mathbf{F}^{sf} - \mu^e \mathbf{E}^{es} = \rho^s \frac{d^s \mathbf{v}^s}{dt}. \quad (3.6)$$

Similarly,

$$\nabla \cdot \boldsymbol{\tau}^f + \rho^f \mathbf{f}^f + \mathbf{F}^{Ef} + \mathbf{F}^{fs} - \mu^e \mathbf{E}^{ef} = \rho^f \frac{d^f \mathbf{v}^f}{dt}, \quad (3.7)$$

$$-\nabla p^e + \rho^e \mathbf{f}^e + \mathbf{F}^{Ee} + \mu^e \mathbf{E}^{es} + \mu^e \mathbf{E}^{ef} = \rho^e \frac{d^e \mathbf{v}^e}{dt}, \quad (3.8)$$

where $d^s/dt$, $d^f/dt$ and $d^e/dt$ are material time derivatives following the solid, fluid and electron continua, respectively. Adding Eqs. (3.6)–(3.8), we obtain the following linear momentum equation for the combined continuum:

$$\nabla \cdot \boldsymbol{\tau} + \rho \mathbf{f} + \mathbf{F}^E = \rho^s \frac{d^s \mathbf{v}^s}{dt} + \rho^f \frac{d^f \mathbf{v}^f}{dt} + \rho^e \frac{d^e \mathbf{v}^e}{dt}, \quad (3.9)$$

where we have denoted the total stress, electric body force and mass density by

$$\boldsymbol{\tau} = \boldsymbol{\tau}^s + \boldsymbol{\tau}^f - p^e \mathbf{1},$$

$$\mathbf{F}^E = \mathbf{F}^{Es} + \mathbf{F}^{Ef} + \mathbf{F}^{Ee} = \mathbf{P}^s \cdot \nabla \mathbf{E} + \mathbf{P}^f \cdot \nabla \mathbf{E} + \mu^e \mathbf{E}, \tag{3.10}$$

$$\rho = \rho^s + \rho^f + \rho^e, \quad \mathbf{f} = \frac{\rho^s \mathbf{f}^s + \rho^f \mathbf{f}^f + \rho^e \mathbf{f}^e}{\rho}. \tag{3.11}$$

The equations in Eqs. (3.6)–(3.8) may also be subtracted from each other to form equations for the differences of various fields.

## 4. Angular Momentum Equations

For the angular momentum equation, the solid continuum needs to be viewed as a combination of a lattice continuum and a bound-charge continuum [7]. The fluid continuum is similar. The electron continuum does not have electric polarization and the related electric body couple. We neglect possible interactive couples among the constituents. Then the separate angular momentum equations for the solid, fluid and electron continua are

$$\varepsilon_{ijk} \tau^s_{jk} + C^{Es}_i = 0, \tag{4.1}$$

$$\varepsilon_{ijk} \tau^f_{jk} + C^{Ef}_i = 0, \tag{4.2}$$

$$\varepsilon_{ijk} \tau^e_{jk} = 0, \tag{4.3}$$

where $\varepsilon_{ijk}$ is the permutation tensor and $\mathbf{C}$ is the electric body couple [7]:

$$C^{Es}_i = \varepsilon_{ijk} P^s_j E_k, \quad C^{Ef}_i = \varepsilon_{ijk} P^f_j E_k, \quad C^{Ee}_i = 0. \tag{4.4}$$

Adding Eqs. (4.1)–(4.3), we obtain

$$\varepsilon_{ijk} \tau_{jk} + \varepsilon_{ijk} P^s_j E_k + \varepsilon_{ijk} P^f_j E_k = 0. \tag{4.5}$$

## 5. Energy Equation

For the energy equation, instead of adding equations for individual constituents, we work on the combined continuum including all three constituents directly. This will prove to be sufficient for deriving the constitutive relations and the heat or dissipation equation later. For the combined continuum, with consideration of the powers of all external loads from mechanical and electrical sources, we have

$$\begin{aligned}
&\frac{\partial}{\partial t} \int_v \left( \frac{1}{2} \rho^s \mathbf{v}^s \cdot \mathbf{v}^s + \frac{1}{2} \rho^f \mathbf{v}^f \cdot \mathbf{v}^f + \frac{1}{2} \rho^e \mathbf{v}^e \cdot \mathbf{v}^e \right) dv + \frac{\partial}{\partial t} \int_v \left( \rho^s \varepsilon^s + \rho^f \varepsilon^f + \mu^e \varepsilon^e \right) dv \\
&= \int_s (\mathbf{t}^s \cdot \mathbf{v}^s + \mathbf{t}^f \cdot \mathbf{v}^f - p^e \mathbf{n} \cdot \mathbf{v}^e - \mathbf{n} \cdot \mathbf{q}) ds \\
&\quad - \int_s \mathbf{n} \cdot \left( \mathbf{v}^s \frac{1}{2} \rho^s \mathbf{v}^s \cdot \mathbf{v}^s + \mathbf{v}^f \frac{1}{2} \rho^f \mathbf{v}^f \cdot \mathbf{v}^f + \mathbf{v}^e \frac{1}{2} \rho^e \mathbf{v}^e \cdot \mathbf{v}^e \right) ds \\
&\quad - \int_s \mathbf{n} \cdot \left( \mathbf{v}^s \rho^s \varepsilon^s + \mathbf{v}^f \rho^f \varepsilon^f + \mathbf{v}^e \mu^e \varepsilon^e \right) ds \\
&\quad + \int_v (\rho^s \mathbf{f}^s \cdot \mathbf{v}^s + \rho^f \mathbf{f}^f \cdot \mathbf{v}^f + \rho^e \mathbf{f}^e \cdot \mathbf{v}^e + \rho r) dv + \int_v W^E dv,
\end{aligned} \tag{5.1}$$

where $\varepsilon$ is the internal energy density. $\mathbf{q}$ is the heat flux. $r$ is the body hear source. We note that the internal energy per unit charge [9] is used for the electron continuum instead of per unit mass so that in the limit of a low-density electron fluid with a negligible mass density we still have its internal energy density. The electric body power $W^E$ has the following expression [7,11]

$$W^E = F_k^{Es} v_k^s + \rho^s E_k \frac{d^s \pi_k^s}{dt} + F_k^{Ef} v_k^f + \rho^f E_k \frac{d^f \pi_k^f}{dt} + E_k J_k', \tag{5.2}$$

$$\boldsymbol{\pi}^s = \mathbf{P}^s / \rho^s, \quad \boldsymbol{\pi}^f = \mathbf{P}^f / \rho^f, \quad \mathbf{J}' = \mu^e (\mathbf{v}^e - \mathbf{v}).$$

The differential form of Eq. (5.1) is

$$\begin{aligned}
\rho^s \frac{d^s \varepsilon^s}{dt} + \rho^f \frac{d^f \varepsilon^f}{dt} + \mu^e \frac{d^e \varepsilon^e}{dt} \\
= \tau_{ij}^s v_{j,i}^s + \tau_{ij}^f v_{j,i}^f - p^e v_{j,j}^e + \rho^s E_k \frac{d^s \pi_k^s}{dt} + \rho^f E_k \frac{d^f \pi_k^f}{dt} \\
- \mathbf{F}^{fs} \cdot (\mathbf{v}^f - \mathbf{v}^s) - \mu^e \mathbf{E}^{es} \cdot (\mathbf{v}^e - \mathbf{v}^s) - \mu^e \mathbf{E}^{ef} \cdot (\mathbf{v}^e - \mathbf{v}^f) + \rho r - q_{i,i}.
\end{aligned} \tag{5.3}$$

Using

$$\frac{\partial \mu^e}{\partial t} + (\mu^e v_k^e)_{,k} = 0, \tag{5.4}$$

we can write Eq. (5.3) as

$$\begin{aligned}
\rho^s \frac{d^s \varepsilon^s}{dt} + \rho^f \frac{d^f \varepsilon^f}{dt} + \mu^e \frac{d^e \varepsilon^e}{dt} - \frac{p^e}{\mu^e} \frac{d^e \mu^e}{dt} \\
= \tau_{ij}^s v_{j,i}^s + \tau_{ij}^f v_{j,i}^f + \rho^s E_k \frac{d^s \pi_k^s}{dt} + \rho^f E_k \frac{d^f \pi_k^f}{dt} \\
- \mathbf{F}^{fs} \cdot (\mathbf{v}^f - \mathbf{v}^s) - \mu^e \mathbf{E}^{es} \cdot (\mathbf{v}^e - \mathbf{v}^s) - \mu^e \mathbf{E}^{ef} \cdot (\mathbf{v}^e - \mathbf{v}^f) + \rho r - q_{i,i}.
\end{aligned} \tag{5.5}$$

## 6. Entropy Inequality

Let $\eta$ be the entropy density. For the entropy inequality or the second-law of thermodynamics of the combined continuum including all three constituents, we have

$$\begin{aligned}
\frac{\partial}{\partial t} \int_v (\rho^s \eta^s + \rho^f \eta^f + \mu^e \eta^e) dv \\
\geq -\int_s (\rho^s \eta^s \mathbf{v}^s \cdot \mathbf{n} + \rho^f \eta^f \mathbf{v}^f \cdot \mathbf{n} + \mu^e \eta^e \mathbf{v}^e \cdot \mathbf{n}) ds + \int_v \frac{\rho r}{\theta} dv - \int_s \frac{\mathbf{q} \cdot \mathbf{n}}{\theta} ds.
\end{aligned} \tag{6.1}$$

Equation (6.1) can be converted to differential form as

$$\rho^s \frac{d^s \eta^s}{dt} + \rho^f \frac{d^f \eta^f}{dt} + \mu^e \frac{d^e \eta^e}{dt} \geq \frac{\rho r}{\theta} - \left(\frac{q_i}{\theta}\right)_{,i}. \tag{6.2}$$

Eliminating $r$ from Eqs. (5.5) and (6.2), we obtain the Clausius−Duhem inequality as

$$\rho^s\left(\theta\frac{d^s\eta^s}{dt}-\frac{d^s\varepsilon^s}{dt}\right)+\rho^f\left(\theta\frac{d^f\eta^f}{dt}-\frac{d^f\varepsilon^f}{dt}\right)+\mu^e\left(\theta\frac{d^e\eta^e}{dt}-\frac{d^e\varepsilon^e}{dt}\right)+\frac{p^e}{\mu^e}\frac{d^e\mu^e}{dt}$$
$$+\tau_{ij}^s v_{j,i}^s+\tau_{ij}^f v_{j,i}^f+\rho^s E_k\frac{d^s\pi_k^s}{dt}+\rho^f E_k\frac{d^f\pi_k^f}{dt} \tag{6.3}$$
$$-\mathbf{F}^{fs}\cdot(\mathbf{v}^f-\mathbf{v}^s)-\mu^e\mathbf{E}^{es}\cdot(\mathbf{v}^e-\mathbf{v}^s)-\mu^e\mathbf{E}^{ef}\cdot(\mathbf{v}^e-\mathbf{v}^f)-\frac{q_i\theta_{,i}}{\theta}\geq 0.$$

The free energy $F$ of the constituents can be introduced through the following Legendre transforms:

$$\begin{aligned}F^s &= \varepsilon^s - \theta\eta^s - E_i\pi_i^s,\\ F^f &= \varepsilon^f - \theta\eta^f - E_i\pi_i^f,\\ F^e &= \varepsilon^e - \theta\eta^e.\end{aligned} \tag{6.4}$$

Then the energy equation in Eq. (5.5) and the Clausius–Duhem inequality in Eq. (6.3) become

$$\rho^s\left(\frac{d^s F^s}{dt}+\eta^s\frac{d^s\theta}{dt}+\theta\frac{d^s\eta^s}{dt}\right)+\rho^f\left(\frac{d^f F^f}{dt}+\eta^f\frac{d^f\theta}{dt}+\theta\frac{d^f\eta^f}{dt}\right)$$
$$+\mu^e\left(\frac{d^e F^e}{dt}+\eta^e\frac{d^e\theta}{dt}+\theta\frac{d^e\eta^e}{dt}\right)-\frac{p^e}{\mu^e}\frac{d^e\mu^e}{dt} \tag{6.5}$$
$$=\tau_{ij}^s v_{j,i}^s+\tau_{ij}^f v_{j,i}^f-P_k^s\frac{d^s E_k}{dt}-P_k^f\frac{d^f E_k}{dt}$$
$$-\mathbf{F}^{fs}\cdot(\mathbf{v}^f-\mathbf{v}^s)-\mu^e\mathbf{E}^{es}\cdot(\mathbf{v}^e-\mathbf{v}^s)-\mu^e\mathbf{E}^{ef}\cdot(\mathbf{v}^e-\mathbf{v}^f)+\rho r-q_{i,i},$$

$$-\rho^s\left(\frac{d^s F^s}{dt}+\eta^s\frac{d^s\theta}{dt}\right)-\rho^f\left(\frac{d^f F^f}{dt}+\eta^f\frac{d^f\theta}{dt}\right)-\mu^e\left(\frac{d^e F^e}{dt}+\eta^e\frac{d^e\theta}{dt}\right)+\frac{p^e}{\mu^e}\frac{d^e\mu^e}{dt}$$
$$+\tau_{ij}^s v_{j,i}^s+\tau_{ij}^f v_{j,i}^f-P_k^s\frac{d^s E_k}{dt}-P_k^f\frac{d^f E_k}{dt} \tag{6.6}$$
$$-\mathbf{F}^{fs}\cdot(\mathbf{v}^f-\mathbf{v}^s)-\mu^e\mathbf{E}^{es}\cdot(\mathbf{v}^e-\mathbf{v}^s)-\mu^e\mathbf{E}^{ef}\cdot(\mathbf{v}^e-\mathbf{v}^f)-\frac{q_i\theta_{,i}}{\theta}\geq 0.$$

## 7. Constitutive Relations

For constitutive relations, we break the stresses and polarizations in the solid and fluid into recoverable and dissipative parts as

$$\begin{aligned}\boldsymbol{\tau}^s &= \boldsymbol{\tau}^{sR}+\boldsymbol{\tau}^{sD},\quad \mathbf{P}^s = \mathbf{P}^{sR}+\mathbf{P}^{sD},\\ \boldsymbol{\tau}^f &= \boldsymbol{\tau}^{fR}+\boldsymbol{\tau}^{fD}=-p^f\mathbf{1}+\boldsymbol{\tau}^{fD},\quad \mathbf{P}^f = \mathbf{P}^{fR}+\mathbf{P}^{fD}.\end{aligned} \tag{7.1}$$

The recoverable constitutive relations satisfy

$$\rho^s \left( \frac{d^s F^s}{dt} + \eta^s \frac{d^s \theta}{dt} \right) + \rho^f \left( \frac{d^f F^f}{dt} + \eta^f \frac{d^f \theta}{dt} \right) + \mu^e \left( \frac{d^e F^e}{dt} + \eta^e \frac{d^e \theta}{dt} \right) - \frac{p^e}{\mu^e} \frac{d^e \mu^e}{dt} \quad (7.2)$$

$$= \tau_{ij}^{sR} v_{j,i}^s + \tau_{ij}^{fR} v_{j,i}^f - P_k^{sR} \frac{d^s E_k}{dt} - P_k^{fR} \frac{d^f E_k}{dt}.$$

Then the energy equation in Eq. (6.5) and the Clausius–Duhem inequality in Eq. (6.6) reduce to

$$\rho^s \theta \frac{d^s \eta^s}{dt} + \rho^f \theta \frac{d^f \eta^f}{dt} + \mu^e \theta \frac{d^e \eta^e}{dt}$$

$$= \tau_{ij}^{sD} v_{j,i}^s + \tau_{ij}^{fD} v_{j,i}^f - P_k^{sD} \frac{d^s E_k}{dt} - P_k^{fD} \frac{d^f E_k}{dt} \quad (7.3)$$

$$- \mathbf{F}^{fs} \cdot (\mathbf{v}^f - \mathbf{v}^s) - \mu^e \mathbf{E}^{es} \cdot (\mathbf{v}^e - \mathbf{v}^s) - \mu^e \mathbf{E}^{ef} \cdot (\mathbf{v}^e - \mathbf{v}^f) + \rho r - q_{i,i},$$

$$\tau_{ij}^{sD} v_{j,i}^s + \tau_{ij}^{fD} v_{j,i}^f - P_k^{sD} \frac{d^s E_k}{dt} - P_k^{fD} \frac{d^f E_k}{dt}$$

$$- \mathbf{F}^{fs} \cdot (\mathbf{v}^f - \mathbf{v}^s) - \mu^e \mathbf{E}^{es} \cdot (\mathbf{v}^e - \mathbf{v}^s) - \mu^e \mathbf{E}^{ef} \cdot (\mathbf{v}^e - \mathbf{v}^f) - \frac{q_i \theta_{,i}}{\theta} \geq 0. \quad (7.4)$$

Equation (7.3) is the heat or dissipation equation. Let the reference coordinates of the solid be $\mathbf{X}$ and the motion of the solid be

$$\mathbf{y} = \mathbf{y}(\mathbf{X}, t) \quad \text{or} \quad \mathbf{X} = \mathbf{X}(\mathbf{y}, t). \quad (7.5)$$

Then

$$\mathbf{v}^s = \left. \frac{\partial \mathbf{y}}{\partial t} \right|_{\mathbf{X}}, \quad v_{j,i}^s = X_{M,i} \frac{d^s}{dt}(y_{j,M}). \quad (7.6)$$

Hence we can write Eq. (7.2) as

$$\rho^s \left( \frac{d^s F^s}{dt} + \eta^s \frac{d^s \theta}{dt} \right) + \rho^f \left( \frac{d^f F^f}{dt} + \eta^f \frac{d^f \theta}{dt} \right) + \mu^e \left( \frac{d^e F^e}{dt} + \eta^e \frac{d^e \theta}{dt} \right) - \frac{p^e}{\mu^e} \frac{d^e \mu^e}{dt} \quad (7.7)$$

$$= \tau_{ij}^{sR} X_{M,i} \frac{d^s}{dt}(y_{j,M}) - p^f v_{j,j}^f - P_k^{sR} \frac{d^s E_k}{dt} - P_k^{fR} \frac{d^f E_k}{dt}.$$

Consider the case when

$$F^s = F^s(y_{j,M}; \mathbf{E}; \theta), \quad F^f = F^f\left(\mathbf{E}; (\rho^f)^{-1}; \theta\right), \quad F^e = F^e(\mu^e; \theta). \quad (7.8)$$

The substitution of Eq. (7.8) into Eq. (7.7) gives

$$\rho^s \frac{\partial F^s}{\partial (y_{j,M})} \frac{d^s(y_{j,M})}{dt} + \rho^s \frac{\partial F^s}{\partial E_k} \frac{d^s E_k}{dt} + \rho^s \frac{\partial F^s}{\partial \theta} \frac{d^s \theta}{dt} + \rho^s \eta^s \frac{d^s \theta}{dt}$$
$$+ \rho^f \frac{\partial F^f}{\partial E_k} \frac{d^f E_k}{dt} + \rho^f \frac{\partial F^f}{\partial (\rho^f)^{-1}} \frac{d^f(\rho^f)^{-1}}{dt} + \rho^f \frac{\partial F^f}{\partial \theta} \frac{d^f \theta}{dt} + \rho^f \eta^f \frac{d^f \theta}{dt}$$
$$+ \mu^e \frac{\partial F^e}{\partial \mu^e} \frac{d^e \mu^e}{dt} + \mu^e \frac{\partial F^e}{\partial \theta} \frac{d^e \theta}{dt} + \mu^e \eta^e \frac{d^e \theta}{dt} - \frac{p^e}{\mu^e} \frac{d^e \mu^e}{dt} \qquad (7.9)$$
$$= \tau_{ij}^{sR} X_{M,i} \frac{d^s}{dt}(y_{j,M}) - p^f v_{j,j}^f - P_k^{sR} \frac{d^s E_k}{dt} - P_k^{fR} \frac{d^f E_k}{dt},$$

which leads to the following recoverable constitutive relations:

$$\tau_{ij}^{sR} X_{M,i} = \rho^s \frac{\partial F^s}{\partial (y_{j,M})}, \quad P_k^{sR} = -\rho^s \frac{\partial F^s}{\partial E_k}, \quad \eta^s = -\frac{\partial F^s}{\partial \theta},$$
$$p^f = -\frac{\partial F^f}{\partial (\rho^f)^{-1}}, \quad P_k^{fR} = -\rho^f \frac{\partial F^f}{\partial E_k}, \quad \eta^f = -\frac{\partial F^f}{\partial \theta}, \qquad (7.10)$$
$$p^e = (\mu^e)^2 \frac{\partial F^e}{\partial \mu^e}, \quad \eta^e = -\frac{\partial F^e}{\partial \theta}.$$

In addition, constitutive relations for the dissipative parts of $\boldsymbol{\tau}^{sD}$, $\boldsymbol{\tau}^{fD}$, $\mathbf{P}^{sD}$ and $\mathbf{P}^{fD}$ are needed. For the interactions among the constituents, while constitutive relations may be given in terms of the interactions represented by $\mathbf{F}^{fs}$, $\mathbf{E}^{es}$ and $\mathbf{E}^{ef}$, in applications the constitutive relations are usually given for $\mathbf{v}^f - \mathbf{v}^s$ which is related to the fluid flux or filtration (Darcy's law), $\mathbf{v}^e - \mathbf{v}^s$ and $\mathbf{v}^e - \mathbf{v}^f$ which are related the electric conduction current (Ohm's law), and $\mathbf{q}$ for the heat flux (Fourier's law). The dissipative constitutive relations are restricted by the Clausius–Duhem inequality in Eq. (7.4). The recoverable and dissipative constitutive relations also need to satisfy the requirements of rotational invariance or objectivity [4].

## 8. Conclusions

In summary, the basic unknown fields are $\rho^s$, $\rho^f$, $\rho^e$, $\varphi$, $\mathbf{v}^s$, $\mathbf{v}^f$, $\mathbf{v}^e$ and the temperature field $\theta$. The basic equations for these fields are the continuity equations in Eqs. (2.1)–(2.3), the charge equation of electrostatics in Eq. (2.8), the linear momentum equations in Eqs. (3.6)–(3.8), and the dissipation equation in Eq. (7.3). There are many other fields but they are related to the above ones by the constitutive relations, charge-to-mass ratio, and kinematic relations, etc. On the boundary surface of a finite body, the tractions or velocities of the constituents, the electrostatic potential or the normal component of the electric displacement (depending on the surface charge/current conditions on the boundary), and the temperature or the normal heat flux may be prescribed. For similar but different problems, the theory needed can be constructed using proper multi-continuum models in a similar way.

# References

[1] O. Coussy, *Poromechanics*, John Wiley & Sons, Chichester, West Sussex, England, 2004.

[2] G.D. Bufalo, L. Placidi and M. Porfiri, A mixture theory framework for modeling the mechanical actuation of ionic polymer metal composites, *Smart Mater. Struct.*, 17, 045010, 2008.

[3] W. Hong, X. Zhao and Z. Suo, Large deformation and electrochemistry of polyelectrolyte gels, *J. Mech. Phys. Solids*, 58, 558–577, 2010.

[4] A.C. Eringen and G.A. Maugin, *Electrodynamics of Continua*, vol. I and vol. II, Springer–Verlag, New York, 1990.

[5] H.F. Tiersten, Coupled magnetomechanical equation for magnetically saturated insulators, *J. Math. Phys.*, 5, 1298–1318, 1964.

[6] H.F. Tiersten, An extension of the London equations of superconductivity, *Physica*, 37, 504-538, 1967.

[7] H.F. Tiersten, On the nonlinear equations of thermoelectroelasticity, *Int. J. Engng Sci.*, 9, 587–604, 1971.

[8] H.F. Tiersten and C.F. Tsai, On the interaction of the electromagnetic field with heat conducting deformable insulators, *J. Math. Phys.*, 13, 361–378, 1972.

[9] H.G. de Lorenzi and H.F. Tiersten, On the interaction of the electromagnetic field with heat conducting deformable semiconductors, *J. Math. Phys.*, 16, 938–957, 1975.

[10] M.G. Ancona and H.F. Tiersten, Fully macroscopic description of bounded semiconductors with an application to the Si-SiO$_2$ interface, *Phys. Rev. B*, 22, 6104–6119, 1980.

[11] H.F. Tiersten, On the interaction of the electromagnetic field with deformable solid continua, in: *Electromagnetomechanical Interactions in Deformable Solids and Structures*, Y. Yamamoto and K. Miya, ed., North-Holland, 1987, pp. 277–284.

[12] M.G. Ancona, Hydrodynamic models of semiconductor electron transport at high fields, *VLSI Design*, 3, 101–114, 1995.

[13] J.S. Yang, *An Introduction to the Theory of Piezoelectricity*, 2nd ed., World Scientific, Singapore, 2018.

[14] J.S. Yang, An alternative derivation of the Landau–Lifshitz–Gilbert equation for saturated ferromagnets, *arXiv*, 2305.18232, 2023.

[15] J.S. Yang, A macroscopic theory of saturated ferromagnetic conductors, *arXiv*, 2306.11525, 2023.

[16] J.S. Yang, A continuum theory of elastic-ferromagnetic conductors, *arXiv*, 2307.16669, 2023.

[17] J.S. Yang, *Theory of Electromagnetoelasticity*, World Scientific, Singapore, 2024.

[18] A.C. Eringen, A mixture theory of electromagnetism and superconductivity, *Int. J. Engng Sci.*, 36, 525–543, 1998.